\documentstyle[psfig,aps]{revtex}

\def\beq{\begin{equation}}
\def\eeq{\end{equation}}
\def\bea{\begin{eqnarray}}
\def\eea{\end{eqnarray}}

\def\omk{\omega _k}
\def\dk{\frac {d^3k}{(2\pi )^3}}
\def\d4k{\frac{d^4K}{(2\pi )^4}}

\tightenlines

\begin{document}
\draft

\title{Quantum Field Theoretic Description of Matter in the Universe}

\author{Markus H. Thoma}

\address{Theory Division, CERN, CH-1211 Geneva 23, Switzerland}

\date{\today}

\maketitle

\begin{abstract}
Quantum field theory at finite temperature and density can be used 
for describing the physics of relativistic plasmas. Such systems are
frequently encountered in astrophysical situations, such as the
early Universe, Supernova explosions, and the interior of neutron stars.
After a brief introduction to thermal field theory the usefulness of this 
approach in astrophysics will be exemplified in three different cases.
First the interaction of neutrinos within a Supernova plasma will be 
discussed. Then the possible presence of quark matter
in a neutron star core and finally the interaction of light with the 
Cosmic Microwave Background will be considered.
\end{abstract}

\section{Introduction}
\label{sec:introduction}

The interaction of elementary particles with a medium in equilibrium or
out of equilibrium can be described 
by applying quantum field theory at finite temperature $T$, chemical
potential $\mu$, or in non-equilibrium.
Using perturbation theory, self energies, scattering amplitudes,
the free energy, etc. can be calculated. In this way  
effective masses, Debye screening, dispersion relations, 
refractive indices, damping, decay and production rates, the equation of 
state and other interesting properties of the interacting particle and the 
medium can be derived. Famous examples are listed below.

\begin{itemize}
\item
Effective in-medium neutrino masses, which lead to the MSW-effect,
can be derived from the neutrino self energy at finite temperature and 
density \cite{Noetzold88}. 
\item
The plasmon effect, which describes the decay of an in-medium
photon (plasmon) in a stellar plasma into a neutrino pair,
$\gamma \rightarrow \nu \bar \nu$, provides an effective
energy loss mechanism for hot and dense stars \cite{Adams63}. 
In a relativistic plasma
the use of thermal field theory increases the emissivity by about a
factor of 3 \cite{Braaten91} compared to previous approaches.
\item
The baryon asymmetry of the Universe can be generated by the sphaleron decay
during the electroweak phase transition. For calculating the sphaleron decay
rate in the early Universe a new method in finite temperature field theory
has been developed \cite{Boedeker98}.   
\end{itemize}

After a brief introduction to some basic concepts of
thermal field theory, three examples of its 
application to astrophysical problems are discussed:
neutrino interactions in a Supernova plasma, strange quark matter in 
neutron stars and as a dark matter candidate, and the interaction of
light with the Cosmic Microwave Background.

\section{Thermal Field theory}
\label{sec:thermal}

Perturbative field theory at finite temperature or density is based either
on the so-called imaginary (ITF) or real time formalism (RTF) (see e.g.
\cite{Thoma00}).
As a simple example for illustrating these methods we will discuss
the $\phi ^4$-theory at finite temperature briefly. The most important 
quantity in perturbative field theory is the Feynman propagator. At zero
temperature it is defined as the vacuum expectation value of the time-ordered
product of two field operators. At finite temperature the vacuum expectation 
value has to be replaced by a thermal expectation value containing a sum over 
all thermally excited states. Using the plane-wave expansion of the fields 
and the canonical ensemble one gets \cite{Thoma00} 
\bea
&& i\> \Delta_F^{T>0} (x-y)=\frac {1}{Z}\> \sum _n \> 
\langle n|{\cal T}\{\phi (x)
\phi (y)\}|n\rangle \> e^{-E_n/T}\nonumber \\
&&{\buildrel x_0>y_0 \over=} \int \dk \> \frac {1}{2\omk }\>
\left \{ [1+n_B(\omk )]\> e^{-iK(x-y)} + n_B(\omk )\>
e^{iK(x-y)}\right \},
\label{propagator}
\eea
where $Z$ is the partition function and 
$n_B(\omega_k)=1/[\exp{(\omega_k/T)}-1]$ the Bose-Einstein 
distribution with $\omega_k^2=k^2+m^2$. Here we use the notation
$K^2=k_0^2-k^2$, $k=|{\bf k}|$. This expression has a simple
physical interpretation. At $T=0$ the propagator describes the creation 
of a scalar particle at the space-time point $y$ (for $x_0>y_0$) and its
propagation to $x$, where it is destroyed. At finite temperature there is 
besides spontaneous creation at $y$, given by the term 1 in the square 
brackets, also induced emission and absorption proportional to $n_B(\omega_k)$
due to the presence of the heat bath.

\begin{figure}
\centerline{\psfig{figure=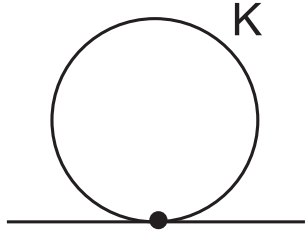,width=4cm}}
\caption{Tadpole diagram}
\label{fig:tadpole}
\end{figure}

In the ITF the propagator given above can be expressed as a sum 
over discrete, imaginary energies (Matsubara frequencies), $k_0=2\pi i nT$,
\beq
i\> \Delta _F^{T>0} (x-y)=iT\> \sum_{k_0} \int \frac{d^3k}{(2\pi )^3}\> 
\frac{i}{K^2-m^2}\>e^{-iK\cdot (x-y)}.
\label{ITF}
\eeq
The ITF will be exemplified by calculating the simplest self energy
in the scalar theory, namely the tadpole diagram shown in 
Figure~\ref{fig:tadpole}.

Modifying the standard Feynman rules by summing over $k_0$ instead of 
integrating we find
\bea
\Pi & = & \frac {i}{2}\> (-i\, 4!\, g^2)\> iT\> \sum _{k_0} \int \dk \>
i\> \frac{1}{K^2-m^2}\nonumber \\
&=&6\> g^2\> \int \dk \> \frac{1}{\omk}\> [1+2n_B(\omk)],
\label{tadpole}
\eea
where $g^2$ is the coupling constant associated with the vertex of the 
$\phi^4$-theory. In the massless case, $m=0$, the integral in 
(\ref{tadpole}) can be done exactly, leading to the simple result
$\Pi=g^2T^2$.

From the self energy we can construct an effective, in-medium propagator
by using the Dyson-Schwinger equation,
$\Delta^*=\Delta+\Delta \Pi \Delta^*$, yielding
\beq
\Delta^*=\frac{1}{K^2-m^2-\Pi}.
\label{effprop}
\eeq
This propagator agrees with the bare one if the bare mass, $m$, 
is replaced by the
effective mass $M=\sqrt{m^2+\Pi}$. The dispersion relation of the scalar field
in the medium follows from the poles of the effective propagator:
$\omega = \sqrt{k^2+M^2}$.

An alternative method to the ITF is the RTF, where the thermal propagator in
momentum space is given by
\beq
i\> \Delta (K) = \frac {i}
{K^2-m^2+i\epsilon } + 2\> \pi \> n_B(|k_0|)\> \delta (K^2-m^2).
\label{RTF}
\eeq
In order to avoid unphysical singularities from products of
$\delta$-functions, which can appear in diagrams with two propagators,
the propagator has been extended to a $2\times 2$-matrix \cite{Landsman87}.
An important advantage of the RTF compared with the ITF is the possible 
generalization to non-equilibrium by replacing the equilibrium distribution
in (\ref{RTF}) by a non-equilibrium.

\begin{figure}
\centerline{\psfig{figure=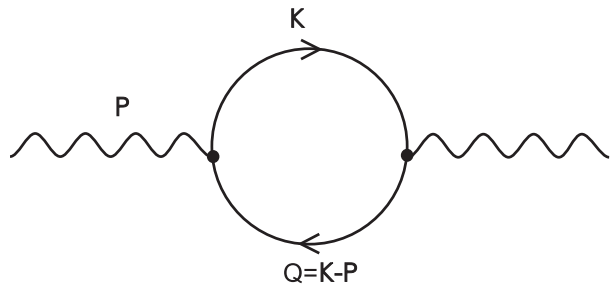,width=8cm}}
\caption{QED Polarization tensor containing an electron-positron loop}
\label{fig:polarization}
\end{figure}

Now we will turn to gauge theories (QED, QCD, electroweak theory)
at finite temperature and density, which are relevant for astrophysical
situations. As an important example we consider the one-loop 
QED polarization tensor in an electron-positron plasma 
shown in Figure~\ref{fig:polarization}.

In contrast to the tadpole diagram the polarization tensor is energy and 
momentum dependent. It can be calculated exactly only in the high-temperature 
limit which is equivalent to the Hard-Thermal-Loop (HTL) limit, in which 
the loop momentum is assumed to be much larger than the external momentum.
The high-temperature limit corresponds to an ultrarelativistic 
electron-positron plasma as it exists in Supernovae.
Due to the breaking of Lorentz invariance by choosing the frame of the
heat bath, the polarization tensor has two independent components, for which 
we take the longitudinal and the transverse. In the HTL limit they read
\bea
\Pi_L(p_0,p)&=&-3\, m_\gamma^2\left (1-\frac{p_0}{2p}\ln 
\frac{p_0+p}{p_0-p} \right ),\nonumber \\
\Pi_T(p_0,p)&=&\frac{3}{2}\, m_\gamma^2\> \frac{p_0^2}{p^2}\> 
\left [1-\left (1-\frac{p^2}{p_0^2}\right )\>
\frac{p_0}{2p}\> \ln \frac{p_0+p}{p_0-p} \right ], 
\label{HTLpolar}
\eea
where $m_\gamma^2=e^2T^2/9$ can be regarded as a thermal photon ``mass''.
However, it should be noted that this mass does not break gauge invariance.

\begin{figure}
\centerline{\psfig{figure=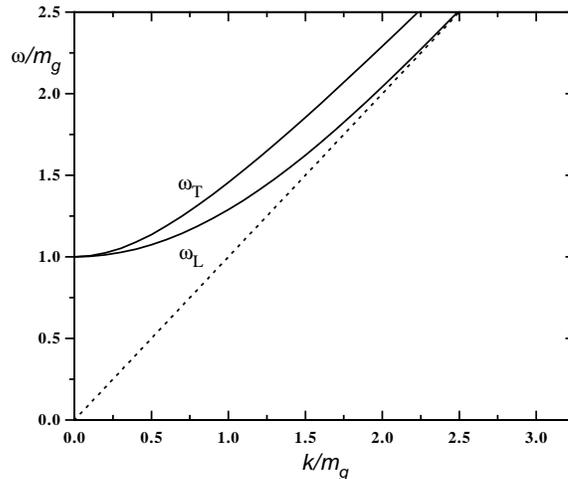,width=10cm}}
\caption{Dispersion relations of longitudinal and transverse photons in
a QED plasma}
\label{fig:dispersion}
\end{figure}

By resumming the polarization tensor using the Dyson-Schwinger equation
we construct an effective 
photon propagator which describes the propagation of collective
photon modes (plasmons) in a QED plasma. The corresponding dispersion 
relations for the longitudinal and the transverse plasma waves are shown in 
Figure~\ref{fig:dispersion}. 

Finally, let me mention that in the case of two-loop self energies, from which 
e.g. damping rates follow, infrared singularities and gauge dependent
results (``plasmon puzzle'') are encountered. Braaten and Pisarski 
\cite{Braaten90}
developed a method, the HTL resummation technique, for avoiding 
these problems, which allows a consistent treatment of gauge theories at 
finite temperature.

\section{Neutrino Interactions in Supernovae}
\label{sec:neutrino}

Neutrinos are copiously emitted from the core of a Supernova. They
interact with the surrounding plasma, providing an effective
mechanism for energy deposition, which pushes the shock wave outwards 
and finally triggers the explosion \cite{Colgate66}. Important 
neutrino interaction processes
in the plasma are $\nu e^\pm \leftrightarrow \nu e^\pm$,
$\bar{\nu} e^\pm \leftrightarrow \bar{\nu} e^\pm$, and
$\nu\bar{\nu}\leftrightarrow e^+ e^-$. 

\begin{figure}
\centerline{\psfig{figure=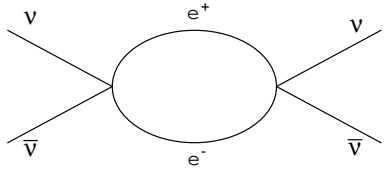,width=5cm}}
\caption{Polarization tensor in Fermi theory with an external neutrino current 
containing an electron-positron loop}
\label{fig:fermi}
\end{figure}

The differential rate for instance for $\nu \bar{\nu}$-annihilation
is given by
\beq
\frac{dR_{\nu \bar \nu}}{d^3q_1d^3q_2} 
=\frac{1}{2q_12q_2} \frac{1}{1-e^{-E/T_e}} M^{\mu \nu}\>  {\rm Im} 
\{\Pi_{\mu \nu}(Q_1+Q_2)\},
\label{diffrate}
\eeq
where 
\beq
M^{\mu\nu}=8(Q_2^\mu Q_1^\nu + Q_1^\mu Q_2^\nu - (Q_1 Q_2) g^{\mu\nu} + i
Q_{1\alpha} Q_{2\beta} \varepsilon^{\alpha\mu\beta\nu})
\label{leptonic}
\eeq
is the leptonic tensor of the neutrino current and $\Pi_{\mu \nu}$
the polarization tensor. To lowest order in the Fermi theory the latter
is given by Figure~\ref{fig:fermi}.

The total rate follows from the differential rate by convolution with 
the neutrino distributions: 
\bea
R_{\nu \bar \nu}=&&\int \frac{d^3q_1}{(2\pi )^3}\int \frac{d^3q_2}{(2\pi )^3}
[f_\nu(q_1)f_{\bar \nu}(q_2)\nonumber \\
&& - (1-f_\nu(q_1))(1-f_{\bar \nu}(q_2))]\> e^{-E/T_e}]\> 
\frac{dR_{\nu \bar \nu}}{d^3q_1d^3q_2} .
\label{totrate}
\eea
Similar expressions for the differential and total rate can be derived for 
$\nu e^\pm$-scattering.

Alternatively the rates can be calculated from the scattering amplitude,
\bea
&&\frac{dR_{\nu \bar \nu}}{d^3q_1d^3q_2} = \frac{1}{2q_12q_2} \int
\frac{d^3p_1}{2E_1(2\pi )^3}[1-n_{e^+}(E_1)] \int \frac{d^3p_2}{2E_2(2\pi)^3}\nonumber \\
&&[1-n_{e^-}(E_2)] \> (2\pi)^4\delta^4(Q_1+Q_2-P_1-P_2)
\sum_{i} \langle |{\cal M}|^2 \rangle ,
\label{amplitude}
\eea
where $n_{e^\pm}(E)=1/[\exp (E\pm \mu)/T+1]$ is the Fermi distribution of
the electrons and positrons in the plasma.

\begin{figure}
\centerline{\psfig{figure=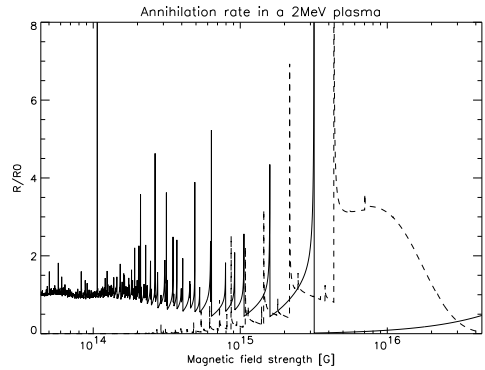,width=7cm}}
\caption{Ratio of the differential rate in a magnetic field to zero
magnetic field for annihilation (solid line) and 
gyromagnetic absorption (dashed line) of a 4 MeV $\nu$ and a
5 MeV $\bar{\nu}$ in an $e^\pm$-plasma at $T=2$ MeV
as a function of $B$}
\label{fig:Bdepend}
\end{figure} 

\begin{figure}
\centerline{\psfig{figure=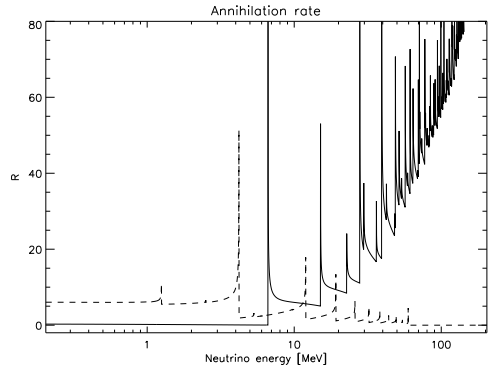,width=7cm}}
\caption{Annihilation and gyromagnetic absorption rate in an 
$e^\pm$-plasma at $T=2$ MeV
as a function of the $\nu$ energy at $B=4 \times 10^{15}$ G}
\label{fig:Edepend}
\end{figure} 

\begin{figure}
\centerline{\psfig{figure=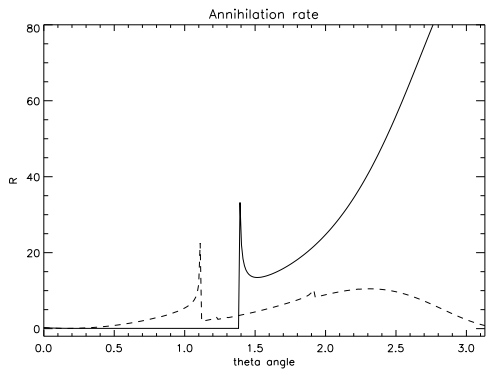,width=7cm}}
\caption{Annihilation and gyromagnetic absorption rate in an 
$e^\pm$-plasma at $T=2$ MeV
as a function of $\theta$, the poloidal angle between the
incoming $\bar{\nu}$ and the magnetic field}
\label{fig:aniso}
\end{figure} 

In addition to medium effects at finite temperature and density we 
consider a strong magnetic field, which plays an important role in 
Supernovae and neutron stars. Recently magnetic field strengths
above $10^{15}$ G have been observed in so-called magnetars \cite{Gotthelf99}.
In a magnetic field the electrons and positrons are in Landau levels,
which leads to a modified electron propagator in the polarization tensor.
The magnetic field can absorb momentum, allowing for new processes
such as the gyromagnetic absorption
$\nu\bar{\nu} e^\pm\leftrightarrow e^\pm$. The 
advantage of starting from the polarization tensor instead of the scattering
amplitude is the automatic inclusion of all possible processes (annihilation,
scattering, gyromagnetic absorption). 

Using the approach described above, we found a strong modification 
of the rates of the various processes in fields at $B>10^{15}$ G. In 
particular there are pronounced peaks in the rates coming from the Landau 
levels, new processes like the gyromagnetic absorption become important, and
the processes show a strong anisotropy, which could play a role in
gamma-ray bursts \cite{Hardy01}. In the 
Figures~\ref{fig:Bdepend} to \ref{fig:aniso} these features are illustrated.

\section{Strange Quark Matter}
\label{sec:strange}

Quark matter consisting of up, down and strange quarks may exist in the 
interior of neutron stars or even in form of strangelets, if quark matter is
stable, i.e., if it has a higher binding energy than iron \cite{Witten84}.
Strangelets have also been discussed as a possible dark matter candidate
\cite{Madsen91}. So far calculations of the equation of state (EoS)
of strange matter are based on an ideal Fermi gas, where sometimes also
one-gluon exchange effects are included \cite{Farhi84}. 

Here we want to 
consider medium effects at finite density and vanishing temperature,
which lead to effective quark masses. Using the one-loop
quark self energy in the high-density limit, an effectice in-medium
quark propagator is constructed. The dispersion relation at zero momentum
following from this propagator defines the effective, density dependent 
quark mass \cite{Schertler97}
\beq
\omega (p=0)=m_q^*=\frac{m_q}{2}+\left 
(\frac{m_q^2}{4}+\frac{g^2\mu^2}{6\pi^2}\right)^{1/2},
\label{qmass}
\eeq
where $\mu $ is the quark chemical potential of the order of 300 MeV.
The coupling constant $g$, typically of the order 2 - 4, can be regarded 
as a parameter describing the medium effect of the effective quark mass, 
which arises naturally due to the interactions as known from mean-field 
theories in many-particle physics. Consequently we replace the ideal Fermi
gas by a quasiparticle gas, where e.g. the particle density is now given by
\beq
\rho(\mu)=\frac{d}{6\pi^2}\> [\mu^2-{m_q^*}^2(\mu )]^{3/2},
\label{density}
\eeq
where $d$ is the degree of freedom, e.g. $d=6$ for strange quarks.
Similar expressions for the energy density and the pressure determining the
EoS can be found.

\begin{figure}
\centerline{\psfig{figure=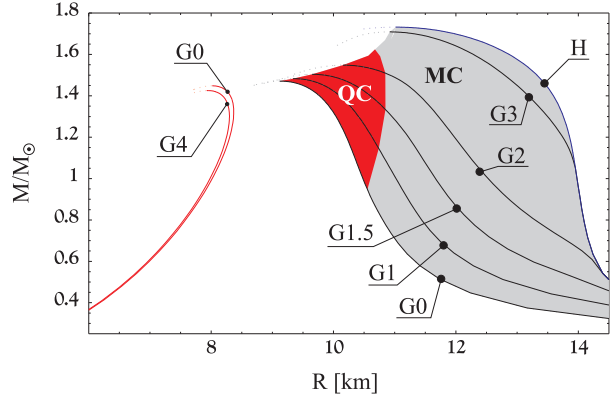,width=8cm}}
\caption{Mass-radius relation of pure strange stars (left) and hybrid
stars (right) for different values of $g$. QC=star has a quark core,
MC=star has a mixed core, H=pure hadron star}
\label{fig:mass_radius}
\end{figure}

The introduction of an effective quark mass, typically of the order of 100 MeV,
reduces the binding energy of strange quark matter and renders the existence
of stable strangelets as a dark matter candidate unlikely \cite{Schertler97}.
Furthermore we studied the influence of the quark medium effects on strange 
stars and hybrid stars, i.e. neutron stars containing a quark matter core,
by solving the Tolman-Oppenheimer-Volkoff equation using the EoS for
quark matter derived above. As shown in Figure~\ref{fig:mass_radius} the
effective
quark mass has a negligible effect on the mass-radius relation of strange 
stars. On the other hand, the presence of a quark matter core, reduces
the radius of a neutron star by typically 20 - 30 \%, which might have
observable consequences \cite{Schertler98,Schertler00}. This reduction is caused by a 
softening of the EoS due to the presence of a mixed phase. It also reduces
the maximum mass of the star to about 1.5 $M_\odot$.

Recently it has been speculated that quark matter at high density should be
in a color superconducting phase \cite{Alford98}. While this could have some 
consequences for the cooling behavior of neutron stars \cite{Blaschke00}, 
it probably does not affect bulk properties, such as the mass-radius relation,
since the pairing takes place only close to the Fermi surface.

\section{Interaction of Light with the Cosmic Microwave Background}
\label{sec:light}

Although the cross section for photon-photon interaction is very small,
light propagating through the Universe interacts continuously
with the Cosmic Microwave Background (CMB).
As discussed in section~\ref{sec:thermal} 
photons may acquire an effective mass 
in a thermal medium. On the other hand, there are very small
upper limits for the photon mass
deduced from laboratory experiments ($m_\gamma <2 \times 10^{-16}$ eV)
and from the galactic magnetic field ($m_\gamma < 10^{-27}$ eV) \cite{Caso98}.
Therefore it is of interest to consider medium effects of light in the CMB. 

\begin{figure}
\centerline{\psfig{figure=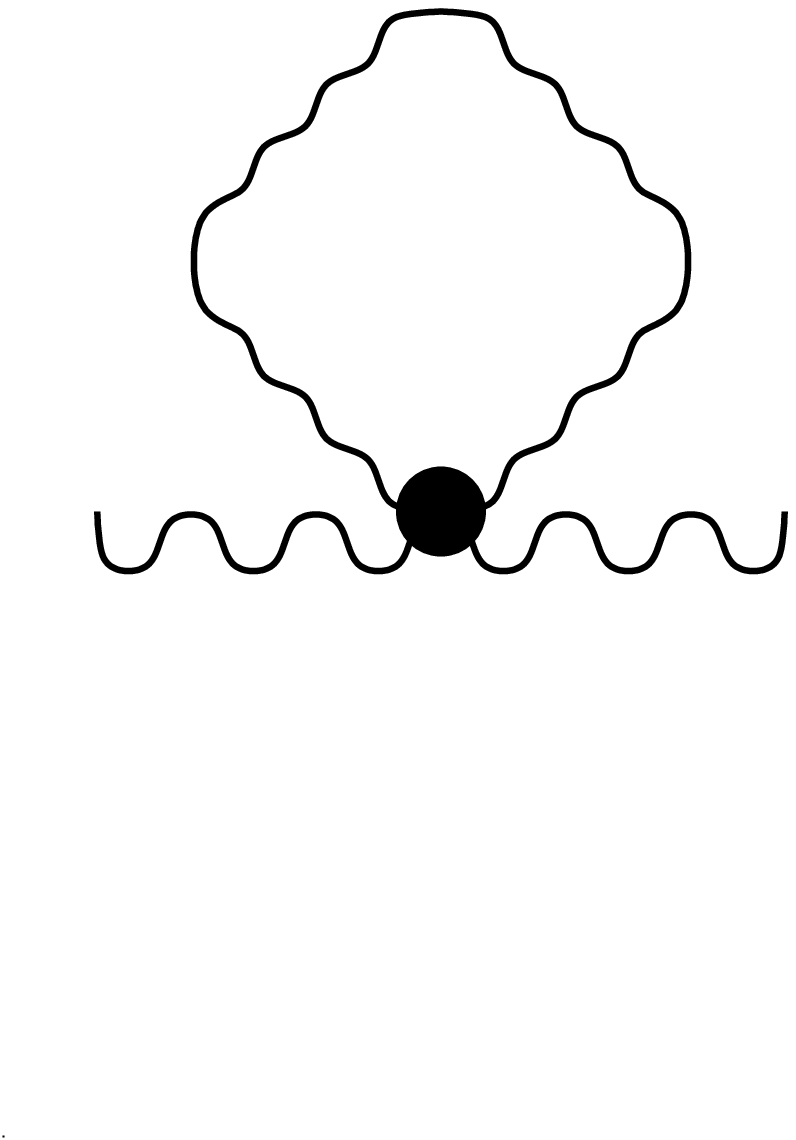,width=4cm}}
\vspace*{-2.5cm}
\caption{Lowest order photon self energy of the effective theory for
photon-photon interaction. The blob denotes the effective 4-photon vertex.}
\label{fig:photon}
\end{figure}

The photon-photon interaction of low energy photons ($E\ll m_e$) can be 
described by an effective Lagrangian \cite{Heisenberg36}
\beq
{\cal L}_I=-\frac{5\alpha^2}{180m_e^4} \> (F_{\mu \nu} F^{\mu \nu})^2
+\frac{7\alpha^2}{90m_e^4}\> F_{\mu \nu} F^{\nu \rho}
F_{\rho \sigma} F^{\sigma \mu},
\label{euler}
\eeq
where $m_e$ is the electron mass, $\alpha $ the fine structure constant,
and $F_{\mu \nu}$ the field strength tensor. The lowest order photon self 
energy following from this Lagrangian is shown in Figure~\ref{fig:photon}.

At finite temperature the longitudinal and transverse photon self energy 
according to Figure~\ref{fig:photon} is given by \cite{Thoma00a}.
\bea
\Pi_L(p_0,p)&=&-\gamma\> p^2, \nonumber \\
\Pi_T(p_0,p)&=&-\gamma\> (p_0^2+p^2),
\label{photself}
\eea
where $\gamma = (44\pi^2/2025)\> \alpha^2\> (T/m_e)^4$.

A gauge invariant definition of the Debye mass of the photon relates it to 
the photon self energy in the following way \cite{Rebhan93}:
$m_D^2-\Pi_{L}(p_0=0,p^2=-m_D^2)=0$. Using (\ref{photself}) one
finds that the Debye mass vanishes, $m_D=0$. Also the effective photon 
``mass'' or plasma frequency, defined as the zero momentum limit of the 
dispersion relation, $\omega (p)=(1-\gamma)\> p$, vanishes,
$\omega(0)=m_\gamma =0$. Hence there is no conflict with the measured 
upper limits of the photon mass.

The electric permittivity and magnetic permeability are also related to
the photon self energy:
\bea
\epsilon &=& 1-\frac{\Pi_L}{p^2}=1+\gamma,\nonumber \\
\frac{1}{\mu}&=&1+\frac{\Pi_T-p_0^2 \Pi_L/p^2}{p^2}=1-\gamma . 
\label{perm}
\eea
From these quantities the phase velocity $v_p$ and the index
of refraction $n$ follow as 
\beq
v_p=\frac{1}{n}=\frac{1}{\sqrt{\mu \epsilon}}=1-\gamma.
\label{phase_velocity}
\eeq
Since the temperature of the Universe drops with time, the phase velocity 
increases continuously to its vacuum value. Whereas it was given by
$v_p=1-5.6\times 10^{-31}$, when radiation and matter decoupled at
about $T=3000$ K, it increased today at $T=2.7$ K to $v_p=1-4.7 \times 
10^{-43}$. Although this is probably not a measurable effect, it is amusing to 
note that the speed of light is not a constant in our Universe.


{}

\end{document}